\documentclass[fleqn,usenatbib]{mnras}

\usepackage{newtxtext,newtxmath}

\usepackage[T1]{fontenc}

\DeclareRobustCommand{\VAN}[3]{#2}
\let\VANthebibliography\thebibliography
\def\thebibliography{\DeclareRobustCommand{\VAN}[3]{##3}\VANthebibliography}

\usepackage{graphicx}	
\usepackage{amsmath}	
\usepackage{orcidlink}
\usepackage{url}
\usepackage{caption}  
\usepackage{multirow}
\usepackage{booktabs}
\usepackage{setspace}
\usepackage{array}   
\usepackage{soul}
\usepackage{ulem}
\usepackage{natbib}

\newcommand{\gwcosmo}{\texttt{gwcosmo} }

\title[The Luminosity of the Darkness]{The Luminosity of the Darkness: Schechter function in dark sirens}

\author[C. Turski et al.]{Cezary Turski,$^{1}$\thanks{E-mail: cezary.turski@ugent.be}
Maria Lisa Brozzetti,$^{2, 3}$\thanks{E-mail: marialisa.brozzetti@pg.infn.it}
Gergely Dálya,$^{4, 5}$
Michele Punturo$^{3}$
and Archisman Ghosh$^{1}$ 
\\
$^{1}$Department of Physics \& Astronomy, Ghent University, Proeftuinstraat 86, 9000 Ghent, Belgium\\
$^{2}$Università degli Studi di Perugia, I-06123 Perugia, Italy\\
$^{3}$Istituto Nazionale di Fisica Nucleare, Sezione di Perugia, Via A. Pascoli, 1, I-06123 Perugia, Italy\\
$^{4}$L2IT, Laboratoire des 2 Infinis - Toulouse, Université de Toulouse, CNRS/IN2P3, UPS, F-31062 Toulouse Cedex 9, France\\
MTA-ELTE Astrophysics Research Group, 1117 Budapest, Hungary\
}

\date{Accepted XXX. Received YYY; in original form ZZZ}

\pubyear{2015}

\begin{document}
\label{firstpage}
\pagerange{\pageref{firstpage}--\pageref{lastpage}}
\maketitle

\begin{abstract}
Gravitational waves (GWs) offer a novel avenue for probing the Universe. One of their exciting applications is the independent measurement of the Hubble constant, $H_0$, using dark standard sirens, which combine GW signals with galaxy catalogues considering that GW events are hosted by galaxies. However, due to the limited reach of telescopes, galaxy catalogues are incomplete at high redshifts. The commonly used GLADE+ is complete only up to redshift $z=0.1$, necessitating a model accounting for the galaxy luminosity distribution accounting for the selection function of galaxies, typically described by the Schechter function. In this paper, we examine the influence of the Schechter function model on dark sirens, focusing on its redshift evolution and its impact on $H_0$ and rate parameters measurements. We find that neglecting the evolution of the Schechter function can influence the prior in redshift on GWs, which has particularly high impact for distant GW events with limited galaxy catalogue support. Moreover, conducting a joint estimation of $H_0$ and the rate parameters, we find that allowing them to vary fixes the bias in $H_0$ but the rate parameter $\gamma$ depends on the evolving Schechter function.
Our results underscore the importance of incorporating an evolving Schechter function to account for changes in galaxy populations over cosmic time, as this impacts rate parameters to which $H_0$ is sensitive.

\end{abstract}




\section{Introduction}

Gravitational wave (GW) astronomy is a novel way of probing the Universe. The most recent GW events catalogue contains 90 binary black holes (BBH), binary neutron stars (BNS) and neutron star - black hole (NSBH) mergers reported by the LIGO-Virgo-KAGRA collaboration (LVK) \citep{LVK2021catalog}. 
Collecting this number of detections allows us to achieve the notable accomplishments in GW cosmology, the measurement of the Hubble constant $H_0$, which yields $H_0=68^{+12}_{-8}~\mathrm{km}\,\mathrm{s}^{-1}\mathrm{Mpc}^{-1}$~\cite{LVK2021_cosmology}.

The concept of measuring $H_0$ with GWs was first described by \cite{Schutz1986}, who proposed leveraging the distance estimates derived from the GW event data in conjunction with the redshift information of the corresponding host galaxies retrieved from galaxy catalogues. In instances where the identification of the host galaxy cannot be determined, the statistical approach can be used, known as the \textit{dark sirens} method. Other studies using GW events, such as spectral sirens \citep{Mastrogiovanni2024} or cross-correlation techniques \citep{Mukherjee2024} yield similar results.
This methodology has the potential to resolve the problem known as Hubble tension: the discrepancy between early-Universe and late-Universe measurements of $H_0$~\citep{Verde_2019_tension}. 
The most precise state-of-the-art measurements yield $H_0=67.36\pm 0.54~\mathrm{km}\,\mathrm{s}^{-1}\mathrm{Mpc}^{-1}$ based on observations of the cosmic microwave background radiation~\citep{Planck2018_cosmo} and $H_0=73.04\pm 1.04 ~\mathrm{km}\,\mathrm{s}^{-1}\mathrm{Mpc}^{-1}$ from type Ia supernovae~\citep{Riess2022}. This results in a $\sim5\sigma$ tension between measurements. Given large uncertainties, the GW measurement of $H_0$ is currently in agreement with both of these measurements. However, improvement of detectors and more GW events will shrink statistical uncertainties in the coming years, therefore it is important to better quantify the systematic errors which will become relatively more important.

At the moment, the main source of systematic uncertainty is the unknown GW population distribution, specifically the merger‑rate and intrinsic mass distribution, as demonstrated in \cite{LVK2021_cosmology}, strongly affects the inferred contributions from galaxies absent from our catalogue part and poorly localized events. The potential bias coming from the unknown population has been accounted for in \cite{Gray2023gwcosmo} where the result has been marginalized over three different GW populations. Future electromagnetic (EM) surveys and enhanced GW detectors will lead to higher galaxy catalogue support and better sky localization of events, and will improve overall accuracy of future $H_0$ measurement. 
The systematic uncertainties arising from the EM aspects of the measurement have not been thoroughly investigated, nevertheless \cite{Turski2023} studied the potential systematics coming from the choice of the galaxy redshift uncertainty model. They found out that this effect is still small compared to the unknown GW population, however its importance is expected to increase with higher support of galaxy catalogues. Moreover \cite{Palmese_2020} studied full photo-$z$ probability density functions (PDFs) versus Gaussian approximations for two best localized GW events: GW170817 (treated as a dark siren) and GW190814. 
Uncertainties arising from peculiar velocities have been studied in \cite{Howlett_2020, Nicolaou_2020, Mukherjee2021}, and these effects are impactful for a handful of nearby events, especially for bright sirens. Peculiar velocities are negligible for further galaxies, and moreover, they are expected to average over a high number of events. 
Moreover, potential bias from weighting the galaxies with luminosity was reported by \cite{Hanselman2024} and \cite{Perna2024}. They found that applying the inconsistent probability for a galaxy to host a GW event based on luminosity may lead to a significant bias in $H_0$.

Currently used by the LVK galaxy catalogue GLADE+ is $90\%$ complete up to 
a redshift $z\,<\,0.08$ in the K band \citep{DalyaGLADE+}, while GW have been observed up to 
a redshift $z\sim 1.2$, GW190403\_051519 being the farthest one \citep{LVK2021catalog}. This creates a vast region where galaxy catalogues do not provide significant support and GWs are still being observed.~Large localization volumes exacerbate the issue. Therefore, to correct for the EM selection bias, the luminosity distribution of the galaxies beyond the reach of telescopes needs to be assessed -- the luminosity of the darkness. Until now, only the constant-in-redshift luminosity function, modelled as Schechter function (SF) \citep{Schechter} is used for the dark standard siren method, however in this paper we consider a luminosity function evolving with redshift for the first time in this context.

Several studies have demonstrated the evolution of the SF, with redshift. In particular, \cite{Lin_1999} demonstrated that luminosity evolution is expected at intermediate redshifts ($0.12 < z <0.5$), and this behaviour is closely linked to the types of galaxy populations, which could have a potential impact on the probability of hosting a compact binary coalescence (CBC) merger.
These elements will be addressed in Section~\ref{subsec:LOS}. As in \cite{Loveday2012MNRAS.420.1239L}, distinct galaxy types were found to correspond to different parameter set in evolving SF, although the characteristic magnitude gets brighter with redshift across all bands examined. Additionally, the comoving number density is lower at high redshifts. Moreover, \cite{Blanton2003ApJ...592..819B} analysed a subset of galaxies from the Sloan Digital Sky Survey (SDSS)~\citep{York2000AJ....120.1579Y} with median redshift $z=0.1$, indicating that a Schechter function with an evolution parameter provides a reasonable fit in the local universe. Strong evidence for evolution came also from \cite{Cool2012}, studying a subset of galaxies from the NOAO Deep Wide-Field Survey~\citep{NOAO2000AAS...197.7709J}.
Whereas \cite{Arnouts2007} examined the evolution of SF in the mid-infrared $K$ band using a spectroscopic and photometric redshift dataset. An evolution of the SF parameter was identified not only locally, but also up to redshifts greater than $z$ = 1.5, as well as a different evolution parameters in value between different galaxy populations.

The rest of this paper is organized as follows. In the Section~\ref{sec:data} we describe the data used in the paper, in the Section \ref{sec:Method} we present our method and the derivation of parameters describing the evolution of Schechter function. In Section \ref{sec:results} we describe the influence of the Schechter function model on the $H_0$ result, and we conclude our findings in Section \ref{sec:conclutions}. Throughout this paper we adopt a flat $\Lambda$CDM cosmology with an $H_0 = 100h = 70~\mathrm{km}\,\mathrm{s}^{-1}\mathrm{Mpc}^{-1}$.

\section{Data}
\label{sec:data}
\subsection{GW data}
\label{subsec:GWdata}
The analysis is performed using the GW events published up to 3rd observing run (O3) by the LVK collaboration, selecting signals with an SNR > 11, as was done in \cite{CosmoGWTC3}: 42 binary black holes (BBHs), 3 neutron star-black hole binaries (NSBHs) and 2 binary neutron stars (BNSs). With the exception of signal GW170817, which was analysed as a \textit{bright siren}, all other signals were treated as \textit{dark sirens}, having no electromagnetic counterpart, which makes it impossible to pin-point the host galaxy. In this work, we used only \textit{dark sirens}, as we only consider the impact of the out-of-catalogue part.\footnote{The corresponding posterior samples and skymaps of events detected up to the first half of the third observing run are available for download at \url{https://zenodo.org/records/6513631}. Data from the second part of the third observing run (O3b) can be accessed at \url{https://zenodo.org/records/5546663}.} In our analysis, we use Madau-Dickinson merger redshift evolution model \citep{MadauDickinson2024} with the low-$z$ slope $\gamma =4.59$, break-point $z_p=2.47$, high-$z$ slope $k=2.86$.  We then release that assumption ing the rate parameters to vary in a joint estimation with Hubble constant in the subsample of 42 BBHs.
For the BH mass model we use a power-law with a Gaussian peak \citep{model2019, model2023} with a slope $\alpha=3.78$, mean of the Gaussian peak $\mu_g=32.27$, width of the Gaussian component $\sigma_g=3.88$, a relative weight between power law and the Gaussian peak $\lambda_{g}=0.03$, a minimal mass of a primary BH $M_{\mathrm{min}}=4.98 M_{\odot}$ and a maximal mass of a primary BH $M_{\mathrm{max}}=112.5 M_{\odot}$.

\subsection{EM data}
\label{subsec:EMdata}
To obtain the galaxy information, we use the most recently released version of the full-sky catalogue known as the Galaxy List for the Advanced Detector Era, GLADE+ \citep{Dalya2018, DalyaGLADE+}. It is the result of cross-matching six different but not independent galaxy catalogues: the 2MASS Photometric Redshift Catalogue \citep{2MPZ_2013}, the 2 Micron All-Sky Survey Extended Source Catalogue \citep{Jarrett2000}, the WISExSCOS Photometric Redshift Catalogue \citep{WISExSCOSPZ}, the Gravitational Wave Galaxy Catalogue \citep{GWGC_2011}, HyperLEDA \citep{HyperLEDA_2014} and the 16th data release of the SloanDigital Sky Survey Data Release 16 \citep{SDSS-DR16Q_2020}.  We use $K_{\mathrm{S}}$-band (denoted $K$-band) in Vega system.

\begin{figure}
    \centering
    \includegraphics[width=0.5\textwidth]{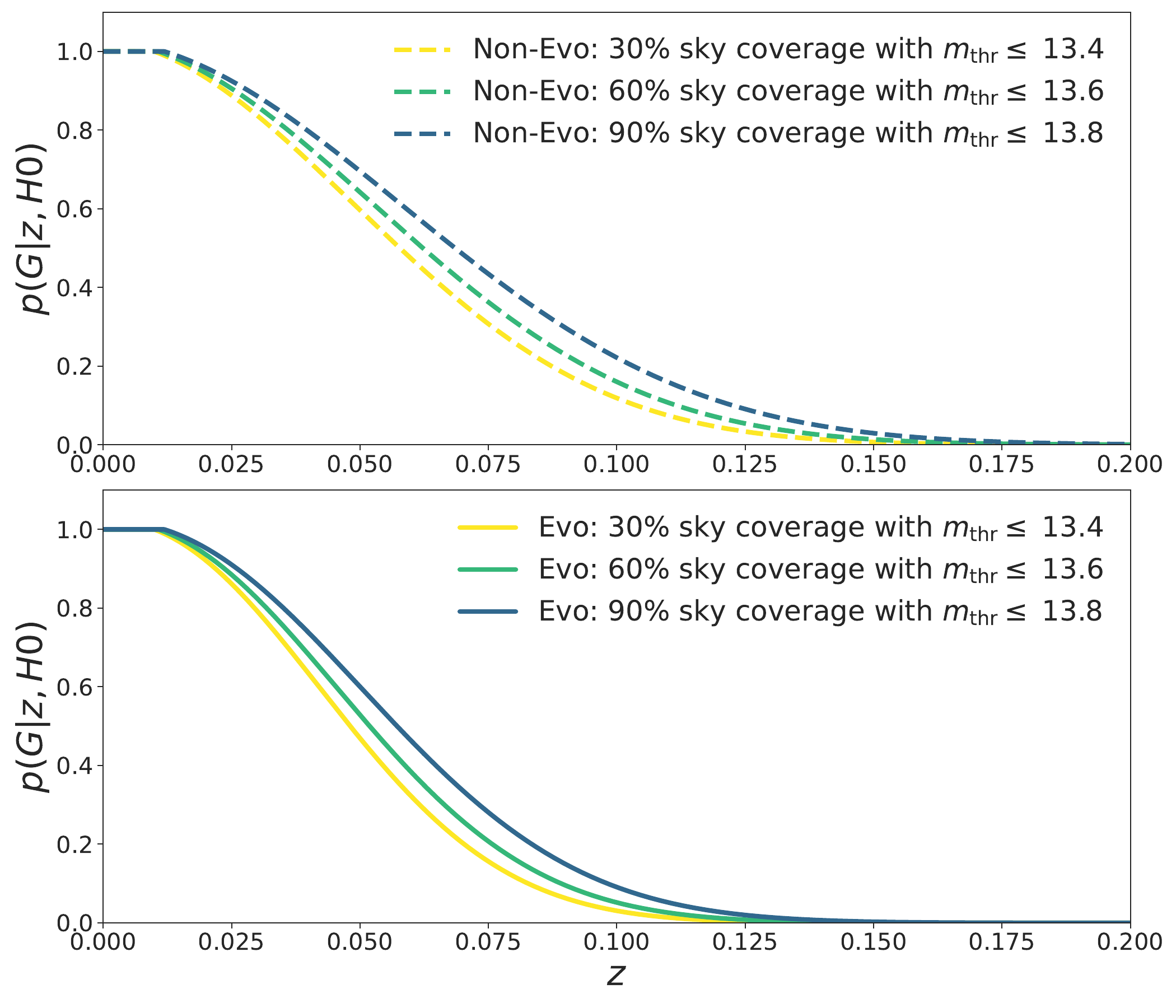}
    \caption{Probability that CBC host galaxies weighted with luminosity in $K$-band are contained in GLADE+, at a given redshift $z$ and $H_0=67.36~\mathrm{km}\,\mathrm{s}^{-1}\mathrm{Mpc}^{-1}$ as in \citet{Planck2018_cosmo}, with constant SF (upper) and evolving SF (bottom).} 
    \label{fig:Kband_Compl}
\end{figure}

The use of a catalogue such as GLADE+ implies that the density of objects varies across the sky, since different surveys cover different sky-areas with different depths. Therefore, in Fig.~\ref{fig:Kband_Compl}, we show the GLADE+ completeness as the percentage of the sky covered at the given magnitude limit, for $K$-band is defined as 
\begin{equation}
    p(G|z, H_0, \Lambda) \, = \, \frac{\int_{M_{\mathrm{min}}(H_0)}^{M_{\mathrm{max}}(z_i,m_{\mathrm{th}}(\Omega_i), H_0)} \Phi(M') dM'}{\int_{M_{\mathrm{min}}(H_0)}^{M_\mathrm{{max}}(H_0)} \Phi(M')dM'}
\end{equation}
assuming that the Schechter function, $\Phi(M)$ is not dependent (top) or dependent (bottom) on the redshift. Where $m_{\mathrm{th}}(\Omega_i)$ contains the information about the survey magnitude limit and the location in the sky. We consider different percentiles of the sky coverage of the catalogue ($30\%,\, 60\%,\, 90\%$), we estimate how the dependency of the SF affects the assessment of missing galaxy, so the completeness of GLADE+ catalogue up to certain median magnitude threshold, $m_{\mathrm{th}}$. Here $5\%$ of the sky has empty pixels, derived by galactic extinction or reddening areas. Both cases indicate that we are missing a lot of EM sources at $z \geq 0.1$. This means that the $H_0$ posterior analysis with events occurring at higher $z$ is dominated by CBC population assumptions. 
However, the galaxy catalogue method is likely to prove more fruitful, given that future surveys will provide deeper catalogues. It will be of considerable interest to observe the forthcoming results with a more complete and comprehensive catalogue, namely UpGLADE (Dálya et al. in preparation).

\section{Method}
\label{sec:Method}
\subsection{Line of sight redshift prior}
\label{subsec:LOS}

We employ the \gwcosmo pipeline \citep{Gray2020gwcosmo,Gray2022gwcosmo,Gray2023gwcosmo}, using the most recent iteration currently accessible. This version incorporates a preliminary step in which the calculation of the line-of-sight (LOS) redshift prior, denoted $p(z)$, is computed, which represents a probabilistic distribution characterizing the likelihood of observing GW events based on the galaxy catalogue. It is given by 
\begin{equation}
    \begin{aligned}
    &p(z|\Omega_i, \Lambda, s, I) = \\&= p(G|\Omega_i, \Lambda, s, I)\iint p(z, M, m|G, \Omega_i, \Lambda, s, I)dMdm  \\&+p(\bar{G}|\Omega_i, \Lambda, s, I)\iint p(z, M, m|\bar{G}, \Omega_i, \Lambda, s, I)dMdm 
    \end{aligned}
\end{equation}
where the first term denotes the in-catalogue part of the calculation, while the second term denotes the out-of-catalogue part, $z$ denotes redshift, $\Lambda$ denotes the cosmological parameters used. The sky is divided into $N_{\mathrm{pix}}$ healpix pixels \citep{Gorski2005} and $\Omega_i$ is the coordinates of a $i$-th pixel. This allows us to apply different magnitude thresholds $m_{\mathrm{th}}$ to different parts of the sky; therefore, the completeness of the catalogue varies and the luminosity function model has a different impact.  $p(G|\Omega_i, \Lambda, s, I)$ $ \big(\text{or } p(\bar{G}|\Omega_i, \Lambda, s, I) \big)$ is a probability that a galaxy hosting a GW event is inside (or outside) the galaxy catalogue. The term $p(z, M, m|\bar{G}, \Omega_i, \Lambda, s, I)$ describes the distribution of redshift and magnitudes of galaxies outside the catalogue range hosting the merger. 
It can be simplified to 
\begin{equation}
\begin{aligned}
    &\iint p(z, M, m|\bar{G}, \Omega_i, \Lambda, s, I) = \\ 
    & = \frac{1}{p(s|\bar{G}, \Omega_i, \Lambda, I)p(\bar{G}|\Omega_i, \Lambda, I)}\\    
    & \quad \times \left[ \int_{M(z, m_{\text{th}}(\Omega_i), \Lambda)}^{M_{\text{max}}(H_0)} p(z, M|\Lambda, I)p(s|z, M, \Lambda, I)dM
    \right] \text{.}
\end{aligned}
\end{equation}
The term $p(s|z, M, \Lambda, I)$ denotes the probability of hosting a CBC given redshift $z$ and luminosity of the host galaxy $M$. This can be used for weighting galaxies with the probability of hosting a merger given its luminosity in some bands denoted as $p(s|M)$. We assume that the probability of a galaxy hosting a GW event is proportional to its stellar mass \citep{Lamberts_2016MNRAS.463L..31L, 2018MNRAS.481.5324M, Artale} and that the luminosity of the galaxy in a $K$-band is a tracer of its stellar mass, whilst the relationship between GW merger probability and galaxy properties, such as star formation rate (SFR) or stellar mass, is currently under investigation and remains poorly constrained. 
The term $p(z, M|\Lambda, I)$ is a distribution of redshifts and magnitudes of galaxies in the Universe. It is usually assumed that galaxies are uniformly distributed in the comoving volume, although  \cite{Dalang2024} looked at completing galaxy catalogues with clustering information. Moreover, we assume that galaxy luminosities are distributed between some minimal and maximal absolute magnitude values ($M_{\text{min}}$ and $M_{\text{max}}$) following the Schechter function distribution $\Phi(z, M)$ that we describe in detail in Section \ref{sec:sch_function}. This leads to 
\begin{equation}
    p(z, M|\Lambda, I) = \frac{\mathrm{d}V}{\mathrm{d}z}(z) \Phi(z, M)
    \label{eq:p(z,M)}
\end{equation}
where $\mathrm{d}V/\mathrm{d}z(z)$ is a comoving volume element. 

\subsection{Schechter function}
\label{sec:sch_function}
\begin{figure}
    \centering
    \includegraphics[width=0.45\textwidth]{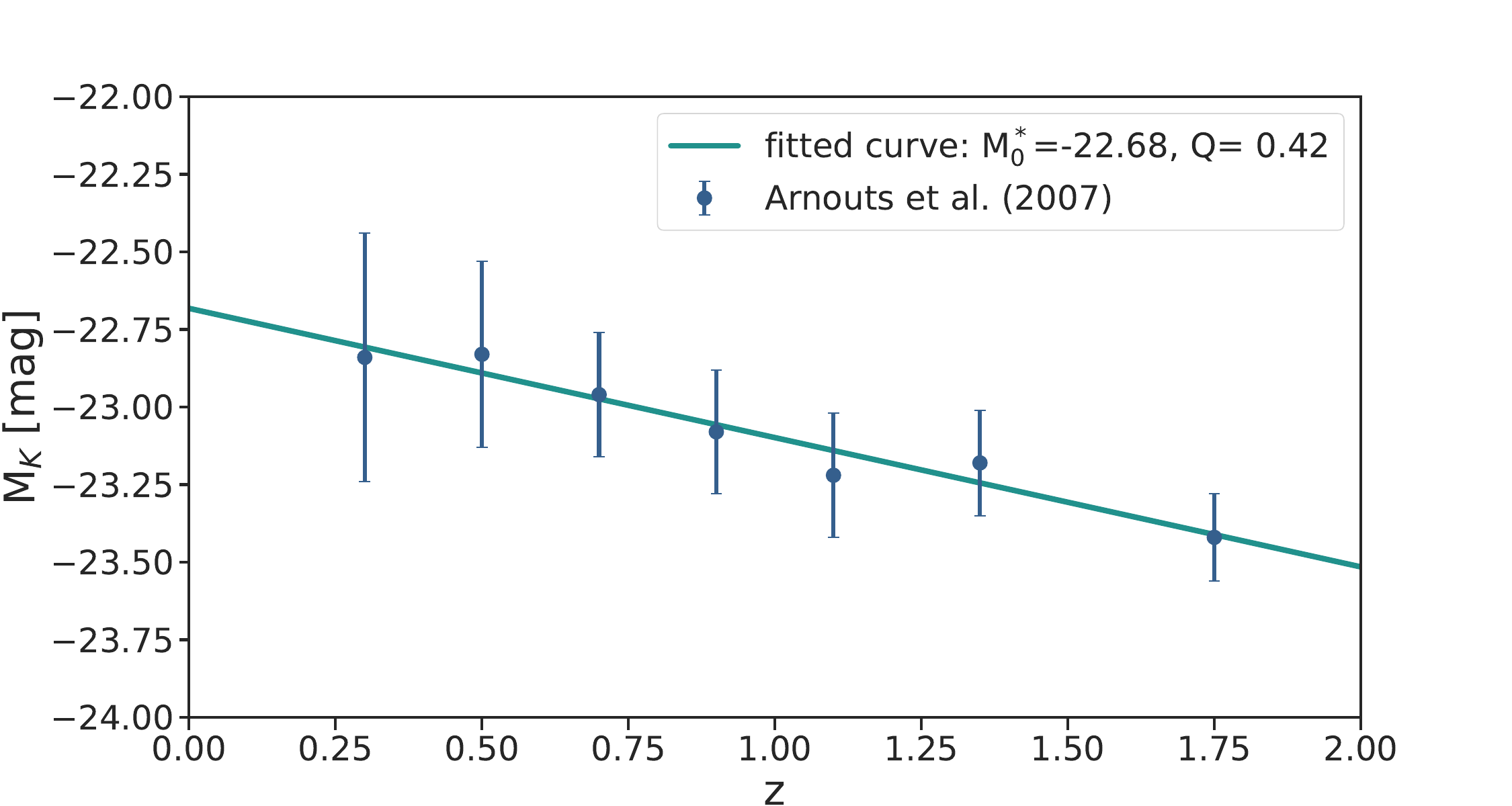}
    \caption{The evolution of the characteristic magnitude $M^*$ with redshift. Blue dots with error bars represent data from \citet{Arnouts2007} and the green line is our fit given by the Eq.~\ref{eq:Mstar} .}
    \label{fig:KbandMstar}
\end{figure}

\begin{figure}
    \centering
    \includegraphics[width=0.45\textwidth]{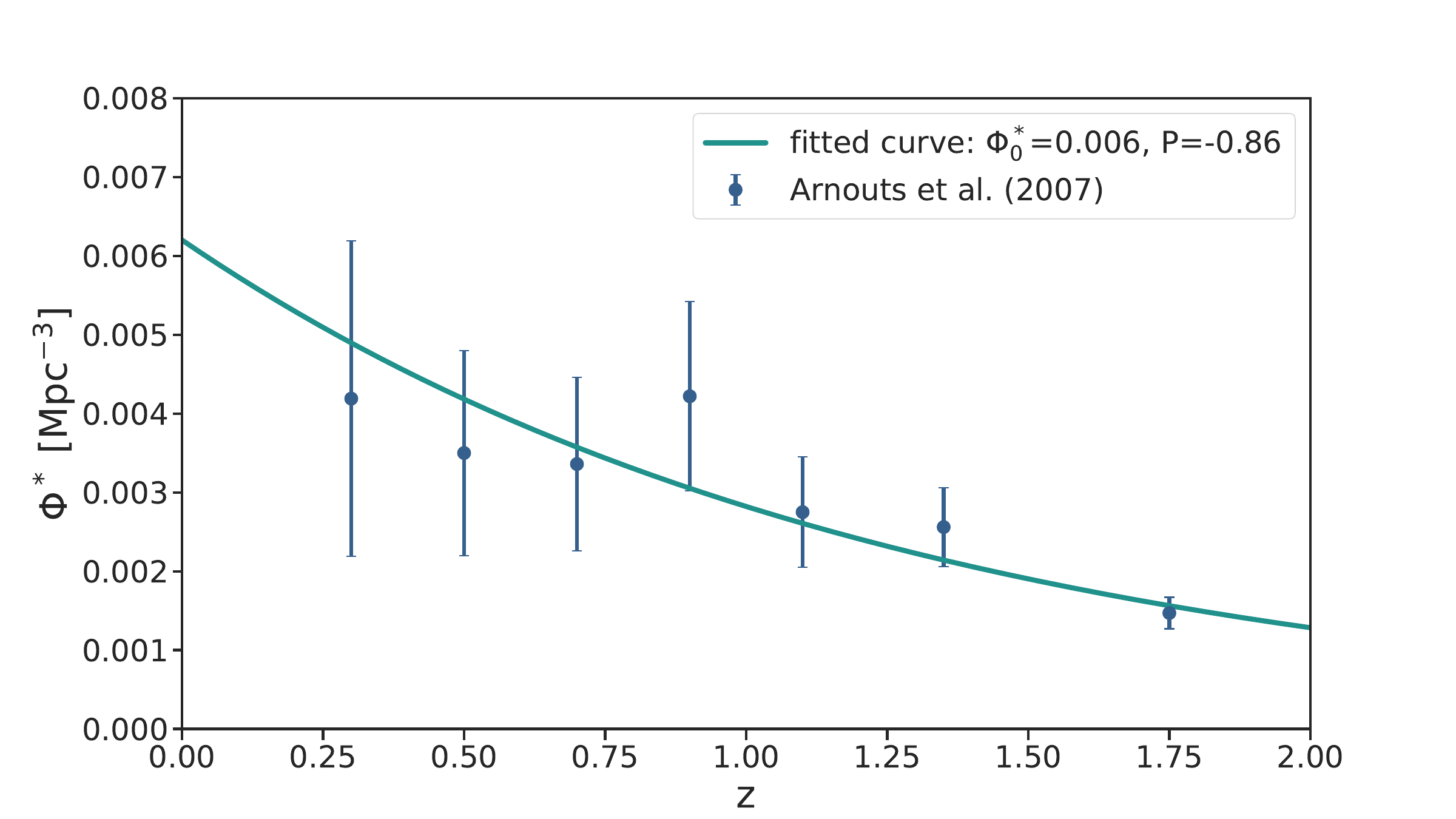}
    \caption{The evolution of the normalization factor $\Phi^*$ with redshift. Blue dots with error bars represent data from \citet{Arnouts2007} and the green line is our fit given by the Eq.~\ref{eq:Phistar}.}
    \label{fig:KbandPhistar}
\end{figure}

The \cite{Schechter} function (SF) model is a commonly used parametrization of the luminosity function (LF) which describes the number density of galaxies as a function of their absolute magnitude $M$, in a unit comoving volume, given by
\begin{equation}
\label{eq:Sch_const}
    \Phi (M) = 0.4\ln 10\,\Phi^*\left(10^{-0.4(M-M^*)}\right)^{1+\alpha}\exp{\left(-10^{0.4(M^*-M)}\right)}
\end{equation}
where $\Phi^*$ is a normalization factor, $M^*$ is a characteristic magnitude at which the luminosity function exhibits a turnover from a power law to a decaying exponential.  The slope $\alpha$ describes  the distribution of fainter galaxies and it has usually negative values. 
Several studies have shown how the LF changes, taking into account different galaxy types, environments and cosmological epochs e.g. in \cite{Lin_1999, Faber2007, Blanton2003ApJ...592..819B, Cool2012, Loveday2012MNRAS.420.1239L}. 
Here we only consider the evolution in redshift of $M^*$ and $\Phi^*$ to study possible significant deviations in the $H_0$ posterior. The evolution of the parameter $\alpha$ is not well studied, since it is the slope of the faint end, it is sensitive to faint galaxies and to galaxy type (See Fig. 11 in \cite{Arnouts2007}), and therefore making it difficult to study its evolution.
It is therefore possible to introduce a redshift parametrization into the equation \ref{eq:Sch_const} to facilitate the study of its dependence: we adopted the formalism introduced by  \cite{Lin_1999} where

\begin{align}
    M^*(z)&= M^*_0 - Qz \label{eq:Mstar}\\
    \Phi^*(z)&=\Phi^*_010^{0.4Pz} \label{eq:Phistar}\\
    \alpha(z)&= \alpha(0)
\end{align}
where $Q$ and $P$ are evolution parameters. To extrapolate the latter parameters we selected values from a previous study of \cite{Arnouts2007}. They used a combined sample of galaxies from Spitzer-SWIRE IRAC \citep{2003Lonsdale}, the VIMOS VLT Deep Survey (VVDS) \citep{2005VIMOS}, UKIDSS \citep{UKIDSS} and the deep Canada France Hawaii Legacy Survey(CFHTLS) \citep{2010CFHTLS} photometric catalogue. Thus both photometric and spectroscopic redshift information are collected up to $z \sim 2$ to investigate the evolution of SF in the $K$-band, together with stellar mass density and different galaxy-type behaviour. The luminosity function is studied using both the Schmidt method \citep{Schmidt1968}, which weights the luminosity contribution according to the maximum volume reached by the survey, and the maximum likelihood method STY from \cite{Sandage1979}. We fixed the value of the faint-end  at $\alpha=-1.1$ as done in \cite{Arnouts2007}. While we fitted values of reported $\Phi^*$ and $M^*$ for $K$-band up to $z=1.5$ for a classical SF and using a least square method. Fig. \ref{fig:KbandMstar} and Fig. \ref{fig:KbandPhistar} show the fitted curves for $M^*$ and $\Phi^*$, respectively, over the chosen redshift range. The obtained values are $\Phi^*_0=0.006 \pm 0.001$ $h^3$Mpc$^{-3}$, $M^*_0=-22.68\pm 0.05$ mag ( corresponding $L^* = 2.68\, \cdot10^{10} L_\odot$), $Q=0.42 \pm 0.04$ and $P=-0.86 \pm 0.13$ with $M_{\text{min}}=-25$ ($L^* = 9.04\, \cdot10^{8} L_\odot$) and $M_{\text{max}}=-19$ ($L^* = 2.27\, \cdot10^{11} L_\odot$) evolving linearly with the same slope as $M^*$, over the selected redshift range.

\begin{figure*}
    \centering
    \includegraphics[width=1.5\columnwidth]{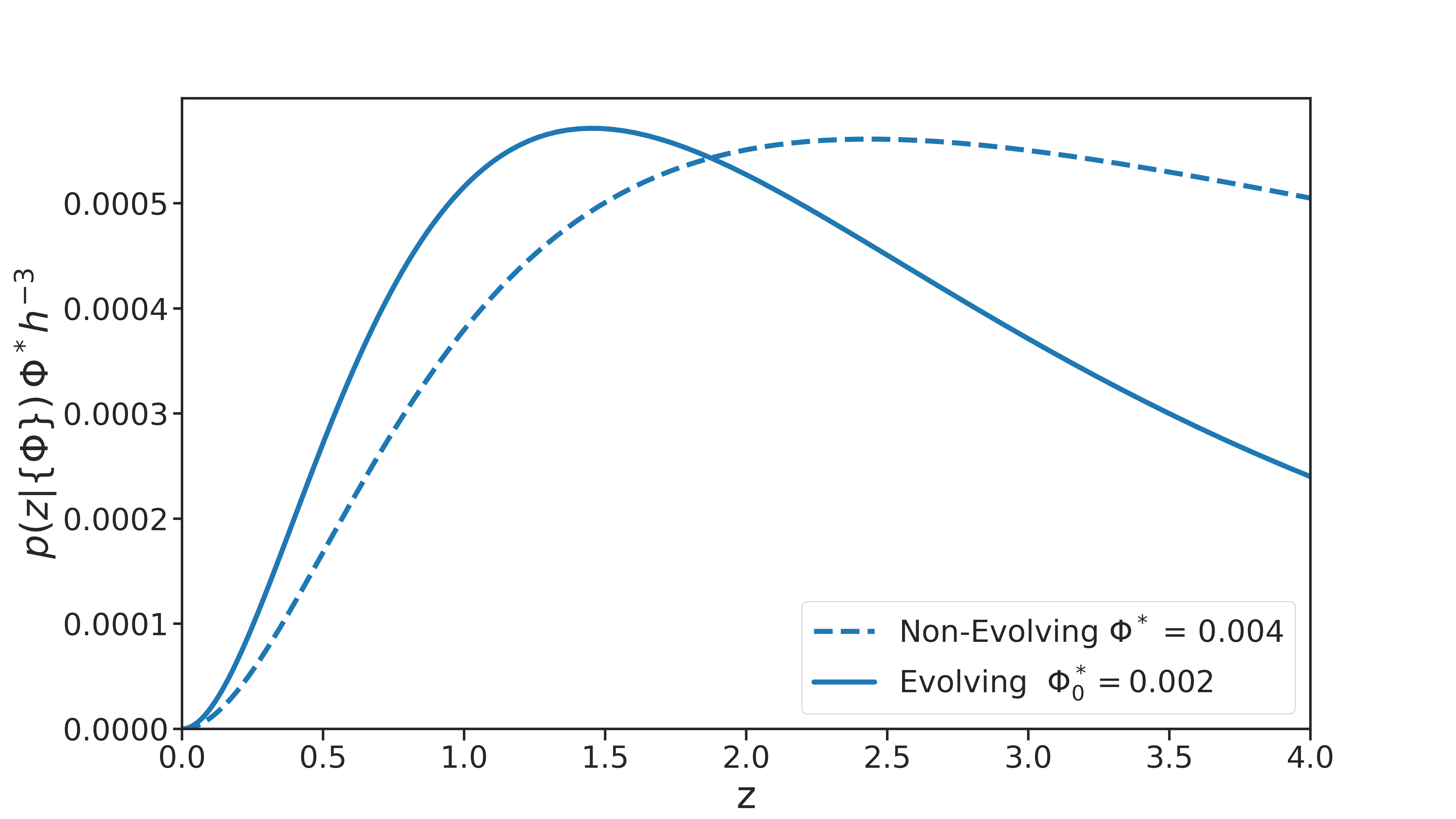}
    \caption{Comparison of the luminosity weighted line-of-sight redshift priors assuming the Schechter function parameters in the $K$-band, multiplied by the  normalisation factor $\Phi^* = 4.0 \times 10^{-2} h^{3}$ Mpc$^{-3}$ for the redshift-independent case (\textit{Non-Evolving}, dashed line), and $\Phi^*_0 = 2.0 \times 10^{-2}  h^{3}$ Mpc$^{-3}$ for redshift-dependent case (\textit{Evolving}, solid line). The Non-Evolving case  remains roughly flat after reaching the maximum between $2 < z <2.5$. Otherwise our work shows a $p(z|\{ \Phi\})$ that increases up to $z\sim 1-1.5$ 
    followed by a fall off at higher redshift.
    }
    \label{fig:LOScomparison}
\end{figure*}

We report the results considering the constant SF, in $K$ band, with $\alpha$= $-1.09$ and $M^*=-23.39$ ($L^* = 5.15\, \cdot10^{10} L_\odot$) from \cite{Kochanek2001ApJ...560..566K}, as in \cite{Gray2023gwcosmo} analysis, however future investigations could include the $B$ band for which GLADE+ is more complete as shown in \cite{DalyaGLADE+} and \cite{brozzetti2024gladenet}. 
Introducing the LF dependency on redshift affects the LOS at higher redshift. In Fig. \ref{fig:LOScomparison} we show the LOS computed from constant and evolving SF multiplied by $\Phi^*\,h^3$ and $\Phi^*_{0}\,h^3$ respectively. The Evolving line remains above the Non-Evolving one up to $z\sim 1.8 $ but the difference is relatively small in the redshift range of detected CBCs. 

\section{Results}
\label{sec:results}

The constant SF LOS was obtained by following the standard procedure with parameters $M^*=-23.39$ ($L^* = 5.15\, \cdot10^{10} L_\odot$) and $\alpha=-1.09$ in $K$ band \citep{Kochanek2001ApJ...560..566K}, as in \cite{Gray2023gwcosmo}. The evolving SF case was derived using the parameters calculated in Section \ref{sec:sch_function}. For the empty catalogue case, the LOS obtained from the set of the Schechter function parameters $\{\Phi\}$ can be obtained by integrating the Eq.~\ref{eq:p(z,M)} with the luminosity weighting $p(s|M)$ over the magnitudes
\begin{equation}
    p(z|\{\Phi\}) \propto \int_{M_{\mathrm{min}}(z)}^{M_{\mathrm{max}}(z)}\frac{\mathrm{d}V}{\mathrm{d}z}(z) \Phi(z, M) p(s|M) \mathrm{d}M \text{.}
\end{equation}
The impact of these assumptions on the LOS is illustrated in Fig. \ref{fig:LOScomparison}. The evolving Schechter function predicts more galaxies at low redshift because it accounts for changes in the galaxy population over time. 
Conversely, the constant SF assumes a constant number density of galaxy per comoving volume, effectively averaging out the redshift evolution. As shown in Fig. \ref{fig:Kband_Compl}, galaxy catalogues appear less complete when interpreted using the evolving SF, as this model emphasizes the higher galaxy number density at low redshifts.
The evolving case remains above the non-evolving one up to $z\sim 1.8 $, showing relatively small difference in the redshift range of detected CBCs. 
At high redshifts, however, the evolving SF predicts fewer galaxies compared to the constant SF. Nonetheless, since the majority of detected GW events originate from redshifts $z < 1$, the LOS at high redshift has minimal impact on the $H_0$ posterior.

\begin{figure}
    \centering
    \includegraphics[width=0.45\textwidth]{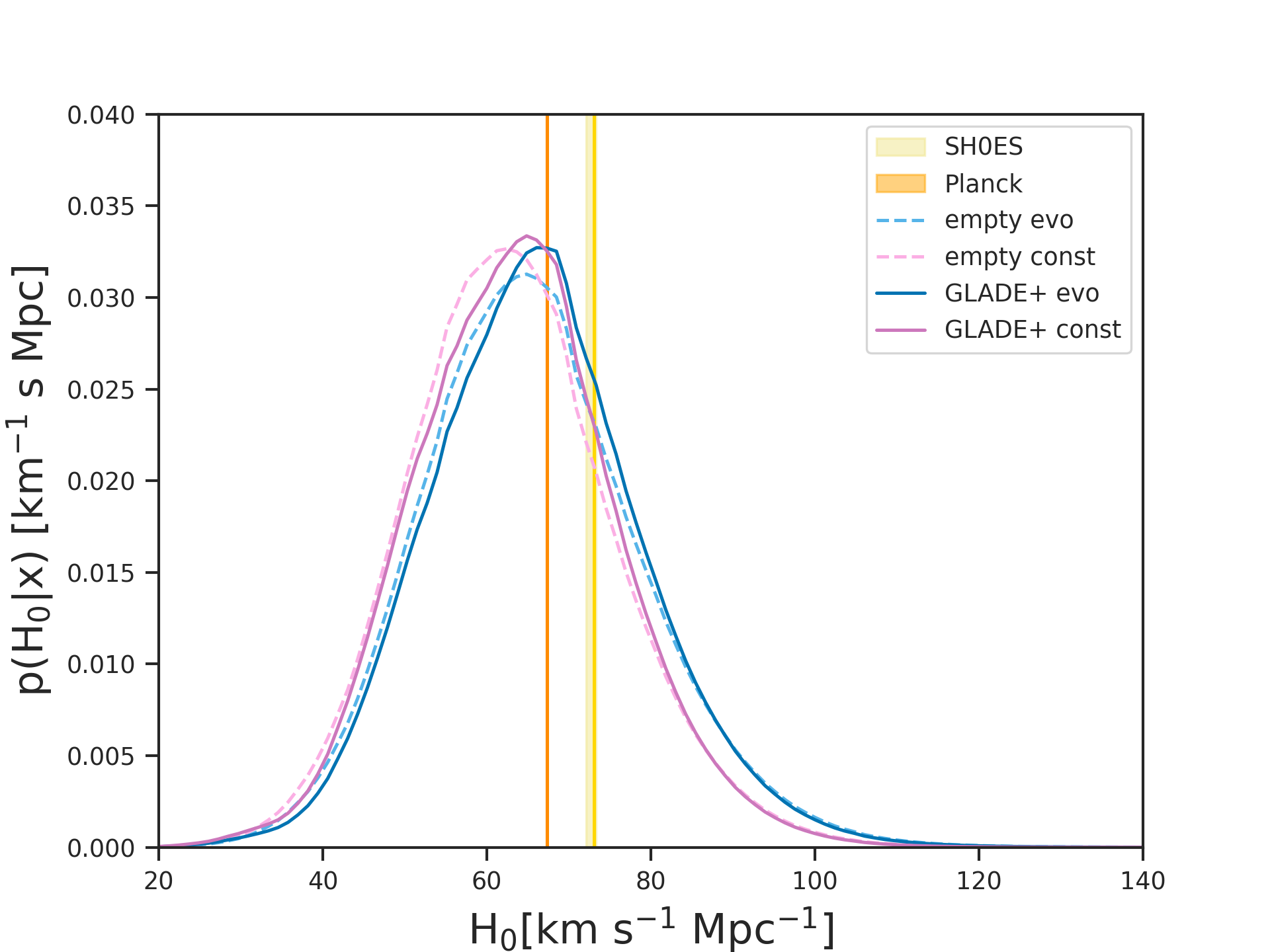}
    \caption{The influence of different Schechter function models on Hubble constant measurement using 46 GW events. Blue (pink) line denotes the evolving (constant) Schechter function case, while solid (dashed) line denotes the usage of GLADE+ (empty catalogue).}
    \label{fig:H0comp}
\end{figure}

\begin{figure}
    \centering
    \includegraphics[width=1.\linewidth]{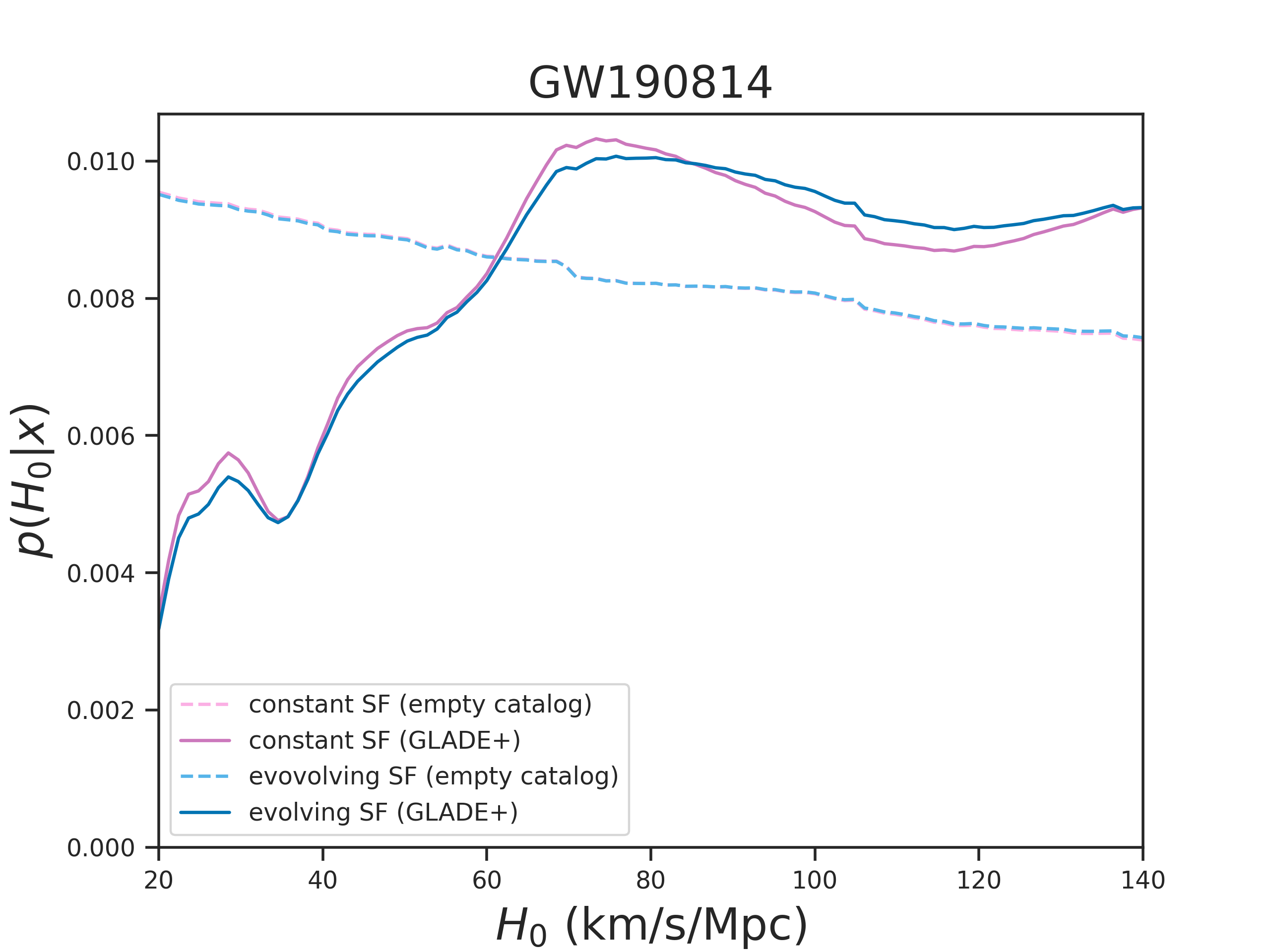}
    \caption{The influence of different Schechter function models on Hubble constant measurement using GW190814 event (luminosity distance $d_L=240^{+40}_{-50}\mathrm{Mpc}$). Blue (pink) line denotes the evolving (constant) Schechter function case, while solid (dashed) line denotes the usage of GLADE+ (empty catalogue).}
    \label{fig:GW190814}
\end{figure}

\begin{figure}
    \centering
    \includegraphics[width=1.\linewidth]{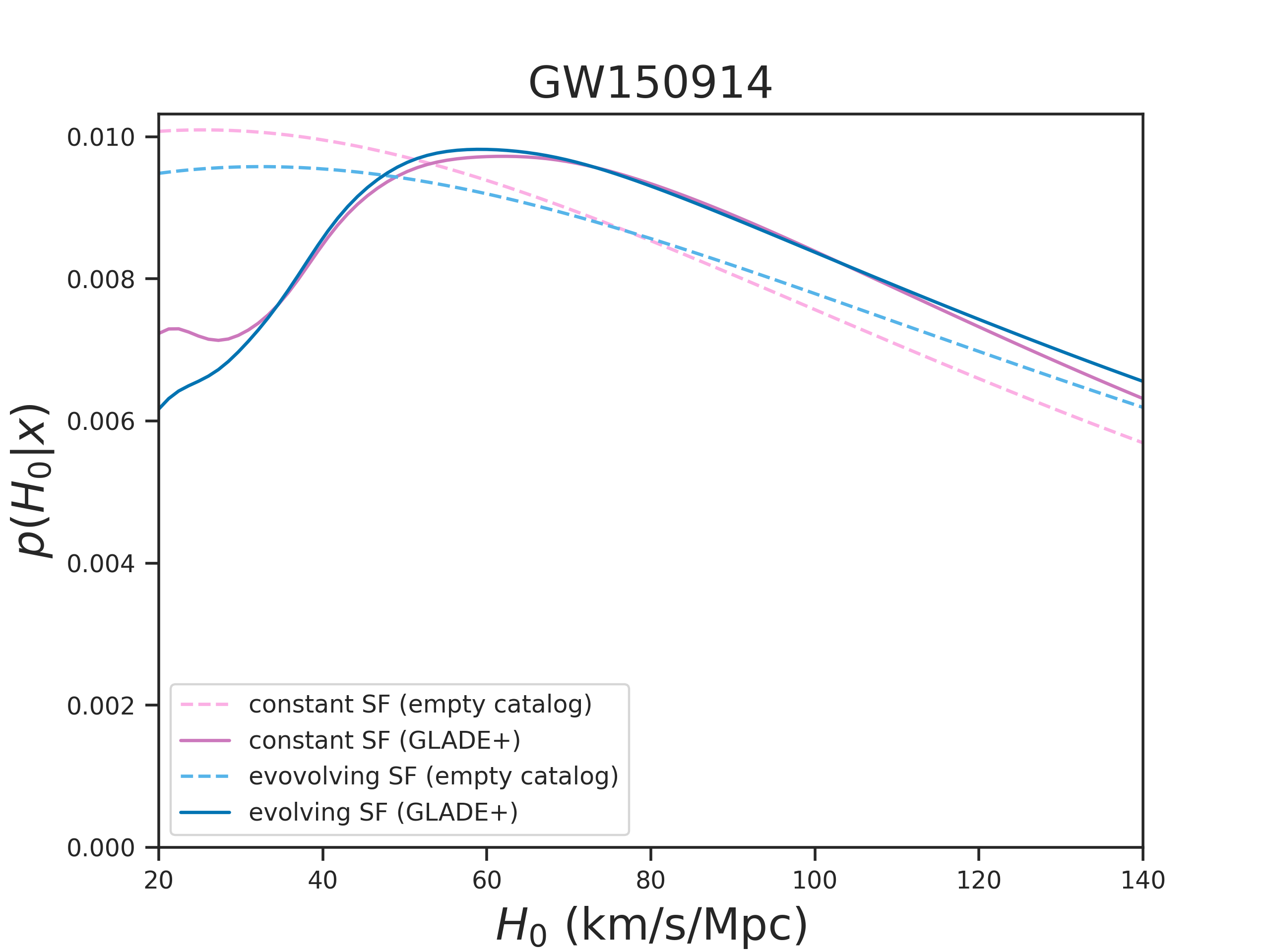}
    \caption{The influence of different Schechter function models on Hubble constant measurement using GW150914 event (luminosity distance $d_L=400^{+100}_{-200}\mathrm{Mpc}$). Blue (pink) line denotes the evolving (constant) Schechter function case, while solid (dashed) line denotes the usage of GLADE+ (empty catalogue).}
    \label{fig:GW150914}
\end{figure}

The impact on the $H_0$ posterior of evolving and constant SF is shown in Fig. \ref{fig:H0comp} and $H_0$ values are reported in Table \ref{table:H0_values}. In the empty catalogue case,  while the comoving galaxy number density decreases with redshift for the evolving Schechter function, the characteristic luminosity increases. This effect biases the constant SF result towards lower values of $H_0$ where luminosities of galaxies remain the same. Using evolving SF corrects for it, yielding slightly higher result, although the shift is well below current statistical uncertainties.
When using a galaxy catalogue, the difference in posteriors is even smaller, as both cases rely on the same underlying luminosity function determined by the galaxy catalogue. This may change in the future once the errors get smaller.
Moreover, the posterior is primarily driven by the GW190814 (Fig. \ref{fig:GW190814}), a well-localized event with strong galaxy catalogue support. The difference in this case arises from the different completeness of the catalogue. For GW190814, the difference between the empty cases is negligible ($O(10^{-5}$)) due to being at low redshift and has relatively small uncertainty on the luminosity distance $d_L=240^{+40}_{-50} \mathrm{Mpc}$, minimizing the impact of luminosity evolution in the evolving SF. However, for other events, the difference is typically larger, as illustrated by GW150914 (Fig. \ref{fig:GW150914}), where constant SF provides higher (lower) support for low (high) values of $H_0$. This illustrates the importance of deep and complete galaxy catalogues as a prior on GW events. 

\begin{figure*}
    \centering
    \includegraphics[width=.8\textwidth]{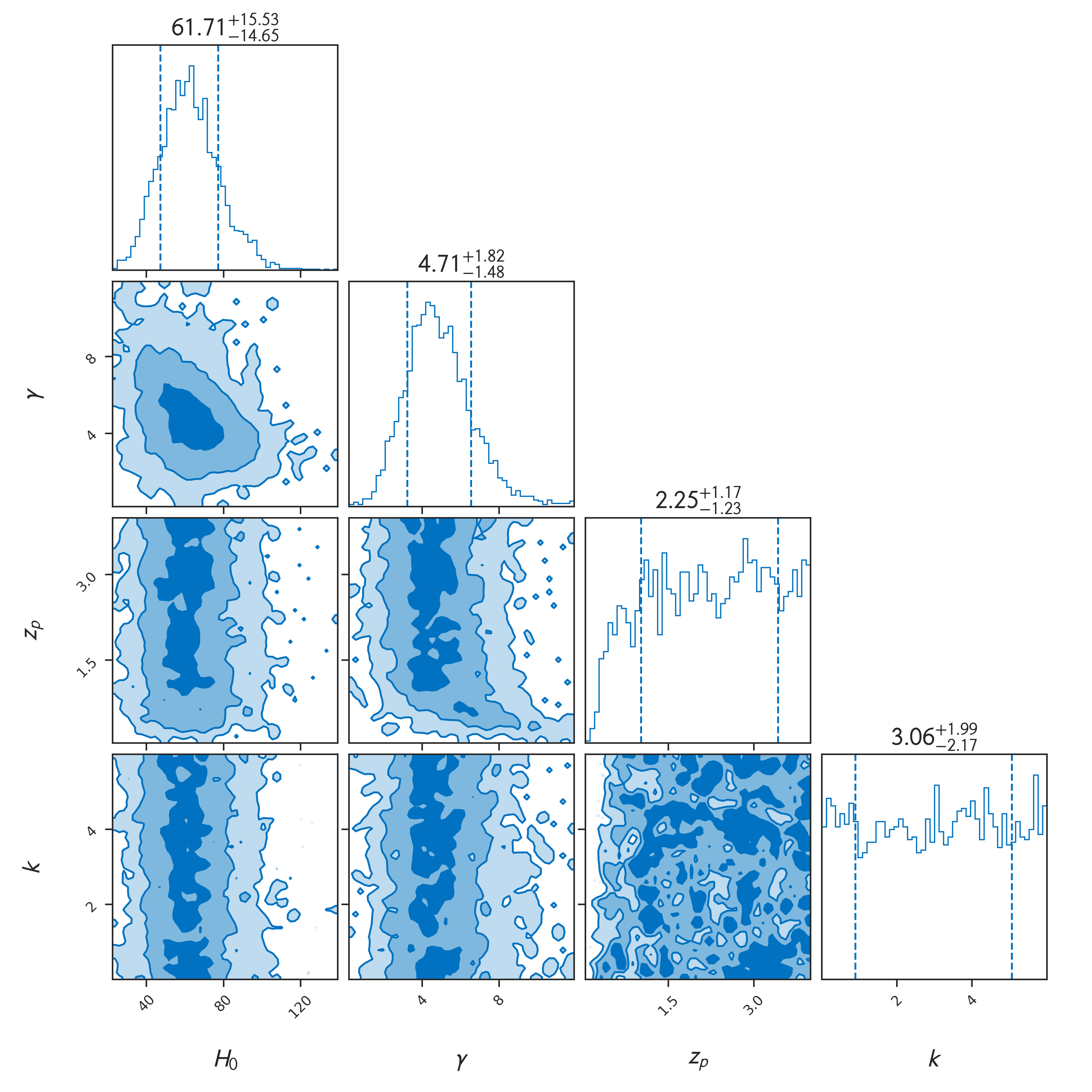}

    \caption{Posteriors on parameters $H_0$, $\gamma$, $z_{\mathrm{p}}$ , $k$ using the 42 BBH events of the
GWTC-3 catalogue, obtained from the population analysis with a \textit{non evolving} Schechter function in gwcosmo pipeline.}
    \label{fig:pop_H0_const}
\end{figure*}

\begin{figure*}
    \centering
    \includegraphics[width=.8\textwidth]{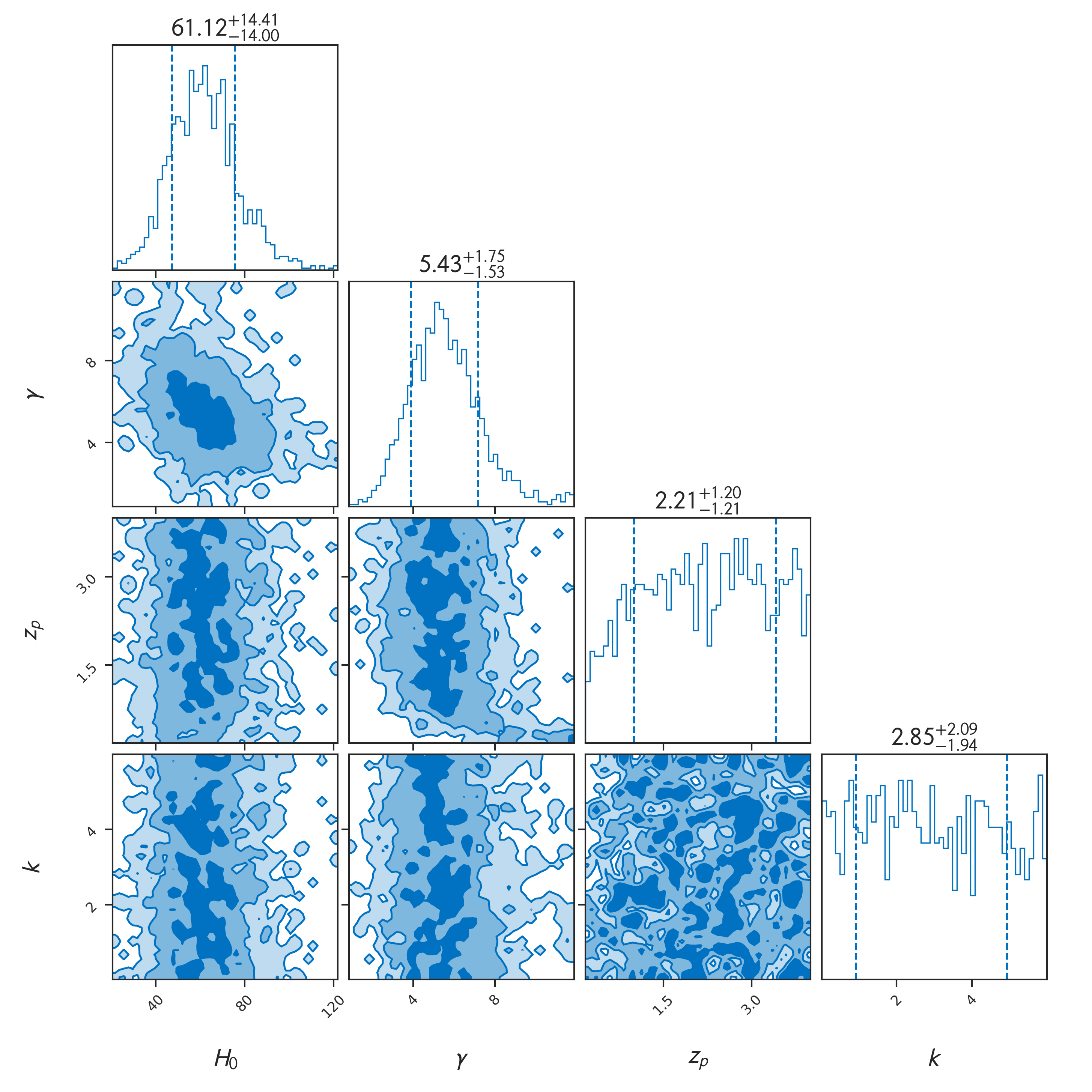}
    \caption{Posteriors on parameters $H_0$, $\gamma$, $z_{\mathrm{p}}$ , $k$ using the 42 BBH events of the
GWTC-3 catalogue, obtained from the population analysis with a \textit{evolving} Schechter function in gwcosmo pipeline.}
    \label{fig:pop_H0_evo}
\end{figure*}

 We then performed an analysis relaxing the assumption about fixed merger rate and allowed the parameters to vary for both constant and evolving SF. The result are respectively in the corner plots in Fig. \ref{fig:pop_H0_const} and Fig. \ref{fig:pop_H0_evo}. Varying of the rate parameters allowed us to mitigate the influence of evolving Schechter function on Hubble constant. However, the rate parameter $\gamma$, which corresponds to the slope of the power-law at low redshift, changes. The Schechter function redshift evolution is degenerate with rate evolution.

\begin{table}
\centering
\begin{tabular}{lcc}
\toprule
\vspace{0.25cm} & Evolving SF & Non-Evolving SF \\
\midrule
\vspace{0.25cm}GLADE+ & $66.06^{+12.32}_{-13.33}$ &  $64.85^{+10.91}_{-14.55}$   \\
\vspace{0.25cm} empty catalogue & $64.85^{+13.33}_{-12.12}$ &  $62.42^{+12.12}_{-12.12}$ \\
\bottomrule
\end{tabular}
\caption{Hubble constant values obtained considering different scenarios. The analyses consist on considering (i) GLADE+ and the evolving SF, (ii) GLADE+ and the non-evolving SF, (iii) and (iv) same SF cases but considering an empty catalogue.}
\label{table:H0_values}
\end{table}

\section{Conclusions}
\label{sec:conclutions}
In this paper, we looked at the impact of evolution with redshift of the Schechter function on Hubble constant measurement using dark sirens. We first constructed the model of the evolution of SF using available data from \cite{Arnouts2007} and compared it with the constant Schechter function model from \cite{Kochanek2001ApJ...560..566K}. We used \textit{dark sirens} GW events with SNR>11. Schechter function models have an impact not only on the out-of-catalogue part of the LOS but they also influence the completeness of the catalogue and thus the in-catalogue part. The result of this has an impact of $H_0$ measurement, see Table \ref{table:H0_values}. This impact is smaller than the impact of the unknown GW population model, which are since then already marginalized over \citep{Gray2023gwcosmo}, but comparable with the influence of the photometric redshift models \citep{Turski2023}, to which it is complementary, as it is the main source of uncertainty in the out-of-catalogue part and the photometric redshift models are the main source of uncertainty of in-catalogue part. 
\begin{table*}
\caption{
}
\label{tab:SF_params_corner}
\centering
\begin{tabular}{lcccc}
\toprule
\vspace{0.15cm} & H$_0$ & $\gamma$  & $z_{\mathrm{p}}$ & k \\
\midrule
\vspace{0.15cm}Non-Evolving SF& $61.71^{+15.53}_{-14.65}$ & $4.71^{+1.82}_{-1.48}$ & $2.25^{+1.17}_{-1.23}$ & $3.06^{+1.99}_{-2.17}$ \\
\vspace{0.15cm}Evolving SF& $61.12^{+14.41}_{-14.00}$ & $5.43^{+1.75}_{-1.53}$ & $2.21^{+1.20}_{-1.21}$ & $2.85^{+2.09}_{-1.94}$ \\
\bottomrule
\end{tabular}
\end{table*}

The uncertainty of the population will decrease as the sensitivity of the detector will increase. Moreover, the mass model will be further restricted with the future planned detectors such as ET \citep{Maggiore_2020JCAP...03..050M,2025_ET_BB} and CE \citep{CE_2022ApJ...931...22S}. This will put more relative importance on the Schechter function, as additionally the detections will come from much greater distances. Moreover, considering the luminosity weighting in the dark siren method, the differences in galaxy population will become important.  
On the contrary, the future electromagnetic surveys such as Vera Rubin Observatory \citep{LSST}, 4-metre Multi-Object Spectroscopic Telescope \citep{4MOST2019} and future data releases from the Sloan Digital Sky Surveys (SDSS) \citep{SDSS} and the Dark Energy Spectroscopic Instrument (DESI) \citep{DESI} will give us deeper galaxy catalogues and thus put more weight on in-catalogue part of the measurement, like the models of redshift profiles, and increase the relative importance of modelling redshifts. 

While the evolution of SF can influence the Hubble constant measurement, its impact can be mitigated by varying the rate parameters. However, it then becomes important to include SF model in rate parameter estimations. 

Furthermore, surveys with higher limiting magnitudes will increase the precision of the SF parameters and their evolution in redshift. Currently, the parameter $\alpha$ is measured only in the very local Universe, and its evolution with redshift remains unknown, however with deeper surveys it will be potentially possible to probe it in a range of redshifts, allowing for modelling the evolution of this parameter as well.


\section*{Acknowledgements}
We would like to thank Freija Beirnaert, Ulyana Dupletsa, Rachel Gray, Simone Mastorgiovanni for the fruitful discussion throughout the project. We are also deeply grateful to Tobia Matcovich and the multi-messenger group at the University of Perugia for their numerous questions and constructive feedback during our meetings. Moreover, we thank Maciej Bilicki for reviewing the manuscript and for many comments during the LVK review. 

The research of CT and AG is supported by Ghent University Special Research Funds (BOF) project BOF/STA/202009/040, the inter-university iBOF project BOF20/IBF/124, and the Fonds Wetenschappelijk Onderzoek (FWO) research project G0A5E24N. They also acknowledge support from the FWO International Research Infrastructure (IRI) grant I002123N and for Virgo collaboration membership and travel to collaboration meetings.

MLB and MP are supported by the Istituto Nazionale di Fisica Nucleare (INFN) for Virgo collaboration and Eistein Telescope membership, and from the project European Union-Next Generation EU, Missione 4 Componente 2, Investimento 3.1 and from PNRR ETIC IR0000004, "EINSTEIN TELESCOPE INFRASTRUCTURE CONSORTIUM" – CUP 53C21000420006".

This material is based upon work supported by NSF's LIGO Laboratory which is a major facility fully funded by the National Science Foundation and Virgo supported by the European Gravitational Observatory (EGO) and its member states. The authors are grateful for computational resources provided by the LIGO Laboratory and supported by National Science Foundation Grants PHY-0757058 and PHY-0823459.

This work makes use of gwcosmo which is available at \hyperlink{https://git.ligo.org/lscsoft/gwcosmo}{https://git.ligo.org/lscsoft/gwcosmo}.

This manuscript was reviewed by the LIGO Scientific Collaboration (document number: LIGO-P2300044) and by the Virgo Collaboration (document number: VIR-0436A-25).

\section*{Data Availability}
The GLADE+ catalogue is available at \hyperlink{https://glade.elte.hu/}{https://glade.elte.hu/}.
All of the GW events we have used in this analysis are available at \hyperlink{https://www.gw-openscience.org/}{https://www.gw-openscience.org/}.



\bibliographystyle{mnras}
\bibliography{bibliography.bib} 





\bsp	
\label{lastpage}
\end{document}